\documentstyle[sprocl]{article}

\input epsf

\def\be{\begin{equation}}
\def\ee{\end{equation}}
\def\bea{\begin{eqnarray}}
\def\eea{\end{eqnarray}}
\def\slashchar#1{\setbox0=\hbox{$#1$}           % set a box for #1
   \dimen0=\wd0                                 % and get its size
   \setbox1=\hbox{/} \dimen1=\wd1               % get size of /
   \ifdim\dimen0>\dimen1                        % #1 is bigger
      \rlap{\hbox to \dimen0{\hfil/\hfil}}      % so center / in box
      #1                                        % and print #1
   \else                                        % / is bigger
      \rlap{\hbox to \dimen1{\hfil$#1$\hfil}}   % so center #1
      /                                         % and print /
   \fi}                                         %

\def\simge{%  ``greater than about'' symbol
    \mathrel{\rlap{\raise 0.511ex
        \hbox{$>$}}{\lower 0.511ex \hbox{$\sim$}}}}
\def\simle{%  ``less than about'' symbol
    \mathrel{\rlap{\raise 0.511ex 
        \hbox{$<$}}{\lower 0.511ex \hbox{$\sim$}}}}

\def\etmiss{\slashchar{E}_T}

\def\tq{{\tilde q}}
\def\tchi{{\tilde\chi}}

\def\lsp{{\tilde\chi_1^0}}
\def\tG{{\tilde G}}
\def\ttau{{\tilde\tau}}
\def\tell{{\tilde\ell}}

\def\GeV{{\rm GeV}}

\def\Meff{M_{\rm eff}}

\def\sgn{\mathop{\rm sgn}}

\def\km{{\rm km}}

\def\fbi{{\rm fb^{-1}}}

\def\mhalf{m_{1/2}}

\catcode`@=11
\def\citenum#1{\csname b@#1\endcsname}
\catcode`@=12

\def\dofig#1#2{\epsfxsize=#1\centerline{\epsfbox{#2}}}
\def\dofigs#1#2#3{\centerline{\epsfxsize=#1\epsfbox{#2}%
   \hfil\epsfxsize=#1\epsfbox{#3}}}
\newdimen\HalfWidth \HalfWidth=\textwidth
\advance\HalfWidth by -0.2in \divide\HalfWidth by 2

\begin{document}

%%%%%%%%%%%%%%%%%%%%%%%%%%%%%%%%%%%%%%%%%%%%%%%%%%
\font\twelvess=cmss10 scaled \magstep1

\begingroup
\parindent=20pt
\thispagestyle{empty}
\vbox to 0pt{
\vskip-1.4in
\moveleft0.75in\vbox to 8.9in{\hsize=6.5in

\centerline{\twelvess BROOKHAVEN NATIONAL LABORATORY}
\vskip6pt
\hrule
\vskip1pt
\hrule
\vskip4pt
\hbox to \hsize{August, 1999 \hfil BNL-HET-99/21}
\vskip3pt
\hrule
\vskip1pt
\hrule
\vskip3pt

\vskip1in
\centerline{\LARGE\bf SUSY MEASUREMENTS AT LHC}
\vskip.5in
\centerline{\bf Frank E. Paige}
\vskip4pt
\centerline{Physics Department}
\centerline{Brookhaven National Laboratory}
\centerline{Upton, NY 11973 USA}

\vskip.75in

\centerline{ABSTRACT}

\vskip8pt
\narrower\narrower
	If SUSY exists at the TeV scale, finding it at the LHC should be
quite easy. The more difficult problem is to separate the various SUSY
channels and to measure masses and other parameters. Substantial
progress on this has been made within the ATLAS and CMS Collaborations.

\vskip1in

	Invited talk at the {\sl International Workshop on Linear
Colliders (LCWS99)} (Sitges, Spain, 28 April -- 5 May 1999).

\vskip0pt

\vfil\footnotesize
	This manuscript has been authored under contract number
DE-AC02-98CH10886 with the U.S. Department of Energy.  Accordingly,
the U.S.  Government retains a non-exclusive, royalty-free license to
publish or reproduce the published form of this contribution, or allow
others to do so, for U.S. Government purposes.

\vskip0pt} % end \vbox to 8.9in
\vss} % end \vbox to 0pt

\newpage
\thispagestyle{empty}
\
\newpage
\endgroup
\setcounter{page}{1}
%%%%%%%%%%%%%%%%%%%%%%%%%%%%%%%%%%%%%%%%%%%%%%%%%%

\title{SUSY MEASUREMENTS AT LHC}

\author{Frank E. Paige}

\address{Brookhaven National Laboratory, Upton, NY 11973, USA}

\maketitle\abstracts{If SUSY exists at the TeV scale, finding it at
the LHC should be quite easy. The more difficult problem is to
separate the various SUSY channels and to measure masses and other
parameters. Substantial progress on this has been made within the
ATLAS and CMS Collaborations.}

\section{Introduction}

	If SUSY exists at TeV scale, it should be quite easy to
discover at the LHC. The main problem is not to discover SUSY but to
make precise measurements of masses and other parameters so as to
understand the SUSY model. Over the last several years, the
ATLAS\,\cite{TDR} and CMS\,\cite{cmssusy} Collaborations have studied
potential to make such measurements for minimal SUGRA models, minimal
GMSB models, and $R$-parity violating models. What can be achieved
depends on the whole pattern of production and decay of SUSY particles
and hence is very model dependent. This paper describes a few of the
many possibilities.

\section{Determination of Masses at SUGRA Point 5}

	``Point 5'' is a minimal SUGRA model with $m_0=100\,\GeV$,
$m_{1/2}=300\,\GeV$, $A_0=300\,\GeV$, $\tan\beta=2.1$, and $\sgn\mu=+$.
It was chosen to give cold dark matter consistent with $\Omega=1$. For
the point there are two characteristic $\tchi_2^0$ decays:  $\tchi_2^0
\to \tell_R^\pm \ell^\mp \to \lsp \ell^+\ell^-$ and $\tchi_2^0 \to \lsp h
\to \lsp b \bar b$.

	The ``effective mass'', $\Meff = \etmiss + p_{T,1} + p_{T,2} +
p_{T,3} + p_{T,4}$, provides a measure of the hardness of the
interaction. The Standard Model background is small with the cuts
$N_{\rm jet}\ge 4$ with $p_T>100,50,50,50\,\GeV$; $\Meff>400\,\GeV$;
and $\etmiss>\max(100\,\GeV,0.2\Meff)$.

\begin{figure}[t]
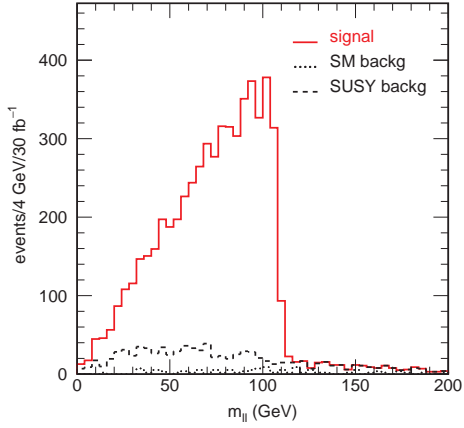

\dofig{\HalfWidth}{p5_mll.epsi}
\caption{Dilepton mass distribution at Point 5. \label{p5_mll}}
\end{figure}

	If events are also required to have two opposite-sign,
same-flavor leptons with ($\ell=e,\mu$) $p_T>10\,\GeV$ and
$|\eta|<2.5$, then the decays $\tchi_2^0 \to \tell_R\ell$ dominate
and produce an endpoint at\,\cite{HPSSY}
$$
M_{\ell\ell}^{\rm max} = M_{\tchi_2^0} 
\sqrt{1 - {M_{\tell_R}^2\over M_{\tchi_2^0}^2}}
\sqrt{1 - {M_{\lsp}^2\over M_{\tell_R}^2}} = 108.93\,\GeV.
$$
This is clearly seen in Figure~\ref{p5_mll}. Forming the combination
$e^+e^-+\mu^+\mu^--e^\pm\mu^\mp$ removes the remaining small
background from two independent decays and allows the endpoint to be
measured to about 0.1\% at high luminosity. The measurement is
sensitive to any $\tilde e_R - \tilde\mu_R$ mass difference at a
similar level.

\begin{figure}[t]
\dofigs{\HalfWidth}{c5_mllq2.epsi}{c5_mllqlow.epsi}
\caption{Left: Smaller of the two $\ell\ell q$ masses at Point~5 and a
fit used to determine the upper limit from squark decay. Right: Larger of
the two $\ell\ell q$ masses with a $\ell\ell$ mass cut and fit used to
determine the lower limit from squark decay. \label{c5_mllq}}
\end{figure}

\begin{figure}[t]
\dofigs{\HalfWidth}{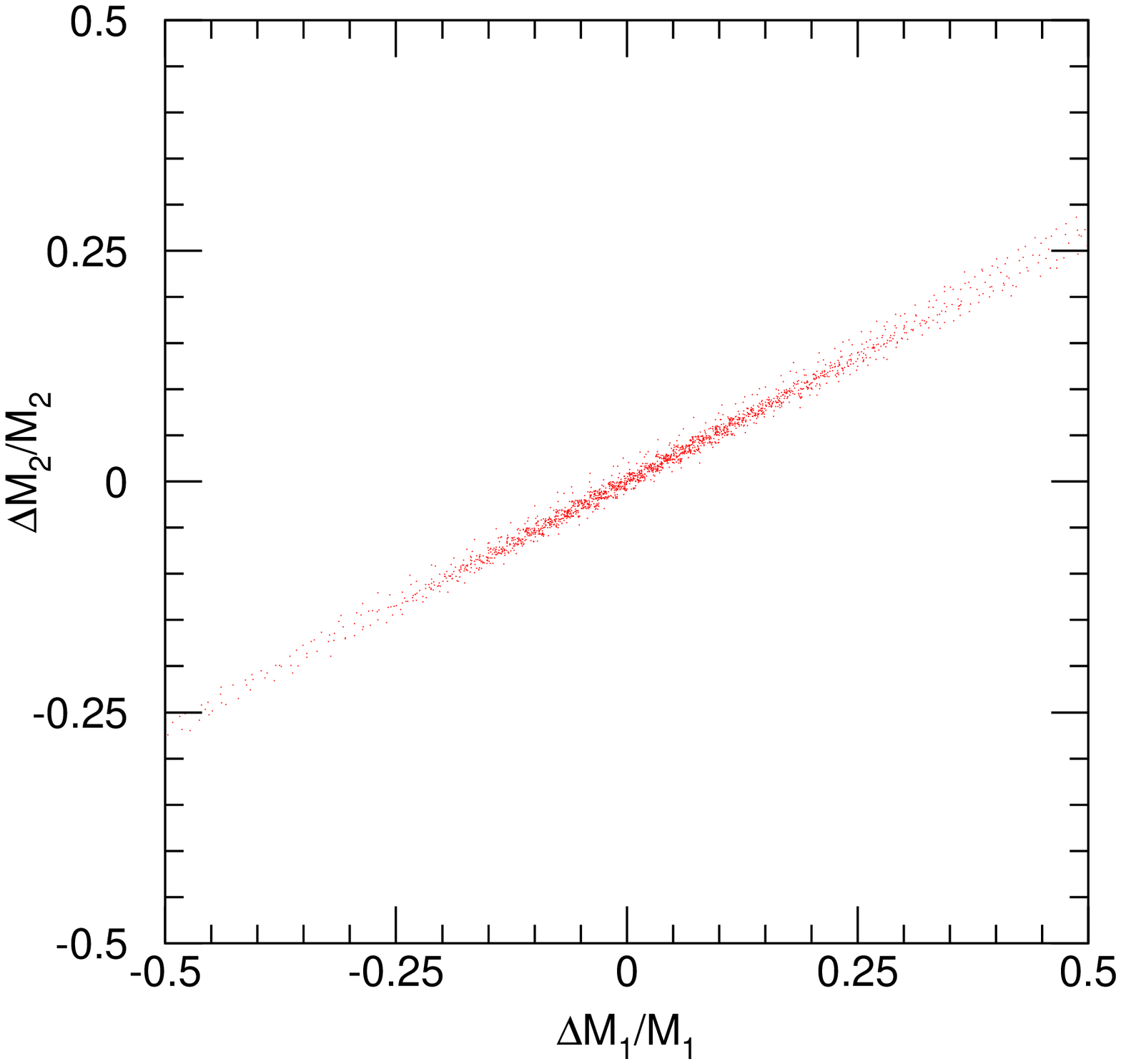}{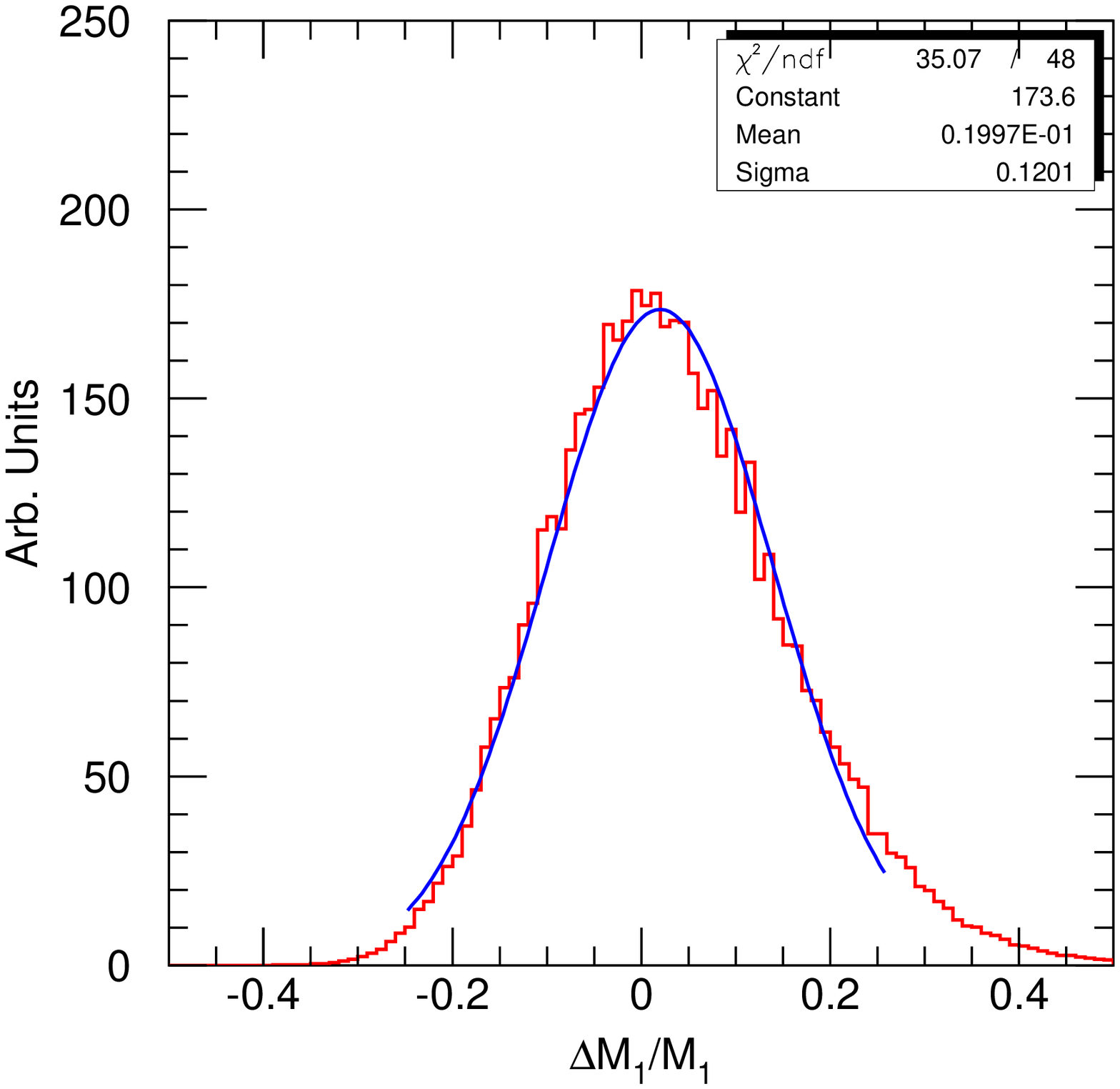}
\caption{Left: Scatter plot of deviations of masses from their nominal
values at Point~5. Right: Projection on the $M_{\lsp}$ axis.\label{c5_m1}}
\end{figure}

	The main source of dileptons is the decay chain $\tq_L \to
\tchi_2^0 q \to \tell_R^\pm \ell^\mp q \to \lsp \ell^+\ell^- q$.
Four-body phase space gives an $\ell\ell q$ endpoint and an $\ell q$
endpoint with functional forms similar to $M_{\ell\ell}^{\rm max}$.
These can be measured by plotting the smaller $\ell\ell q$ mass using
the two hardest jets, and then by plotting the $\ell q$ mass for events
with only one $\ell\ell q$ mass below $600\,\GeV$. The $\ell\ell q$ mass
distribution is shown in Figure~\ref{c5_mllq} together with a fit used
to extract the position of the edge. While the resolutions is only
$\sim10\%$, it is possible to fit these edges with good precision.
However, there are not enough constraints to determine all the masses
involved. If a lower limit is set on the $\ell\ell$ mass, $M_{\ell\ell}
> M_{\ell\ell}^{\rm max}/\sqrt{2}$, then there is also a lower limit on
the $\ell\ell q$ mass. This can be measured by plotting the larger of
the two masses formed by the $\ell\ell$ pair and one of the two hardest
jets; see Figure~\ref{c5_mllq}. A fit with the resolution constrained to
that obtained in the other jet measurements gives $M_{\ell\ell q}^{\rm
min} = 283.7^{+4.4}_{-4.5}\,\GeV$.

	By combining all these measurements the $\tq_L$, $\tchi_2^0$,
$\tell_R$, and $\lsp$ masses can all be determined using only
kinematics.\cite{TDR} The masses are highly correlated, as can be seen
in the scatter plot in Figure~\ref{c5_m1}. The projection of this
scatter plot, also shown in Figure~\ref{c5_m1}, determines the $\lsp$
mass to about 12\%, assuming a 2\% error on the $\ell\ell q$ lower edge.
A fit of these measurements to the minimal SUGRA model gives very small
errors:  $m_0=100.0 \pm 1.4\,\GeV$ (1.4\%), $m_{1/2}=300.0\pm 2.7\,\GeV$
(0.9\%), $\tan\beta=2.00 \pm 0.11$ (5.5\%), and $\mu=+1$. However, $A_0$
is not constrained --- the weak scale $A_t$ is insensitive to it. Also,
if one allows the $\underline{5}$ and $\underline{10}$ scalar masses to
be different at the GUT scale, then the $\underline{5}$ mass is also
poorly constrained: it is hard to measure $\ell_L$ at the LHC. This is
an obvious place where a future linear collider could contribute.

\section{\boldmath $\tau$ Signatures for large $\tan\beta$}

	For $\tan\beta\simle10$ usually have at least one of
$\tchi_2^0 \to \lsp\ell^+\ell^-$, $\tell^\pm\ell^\mp,$ or $\lsp h$ is
generally available. But for $\tan\beta\gg1$ the only allowed two-body
decay, and hence the dominant decay, may be $\tchi_2^0 \to
\ttau_1^\pm\tau^\mp \to \lsp\tau^+\tau^-$. ``Point 6'' is a minimal
SUGRA point with $m_0=200\,\GeV$, $\mhalf=200\,\GeV$, $A_0=0$,
$\tan\beta=45$, and $\sgn\mu=-$ for which this is the case. 

\begin{figure}[t]
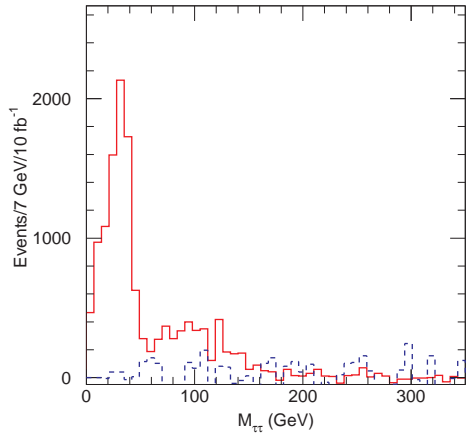

\dofig{\HalfWidth}{p6subtracted.epsi}
\caption{Visible $\tau\tau$ mass distribution at Point 6 after cuts.
\label{p6subtracted}}
\end{figure}

	A clean sample of SUSY events can be selected with multi-jet,
$\Meff$, and $\etmiss$ like those above. An algorithm to reconstruct
$\tau$ hadronic decays with tracking and electromagnetic calorimetry
was developed using the full simulation of ATLAS.\cite{TDR} This
algorithm is biased towards high-mass decays to optimize the
$\tau\tau$ mass measurement rather than $\tau$ identification.  It
gives $\langle M_{\rm visible}\rangle = 0.66M_{\tau\tau}$ with
$(\sigma/M)_{\rm visible} = 0.12$. The rejection factor for light jets
is about 15. The $\tau^+\tau^- - \tau^\pm\tau^\pm$ mass distribution is
shown in Figure~\ref{p6subtracted}. If the $\tau$'s could be measured
perfectly, there would be a sharp edge like that in
Figure~\ref{p5_mll} at $59.64\,\GeV$. While the edge is smeared, a
clear structure remains.  The position of this edge can be determined
to an estimated $3\,\GeV$ for $10\,\fbi$. This can then be used as a
starting point for further partial reconstructions.\cite{TDR} A case
like this one is clearly more difficult and should be studied for
linear colliders.

\begin{figure}[t]
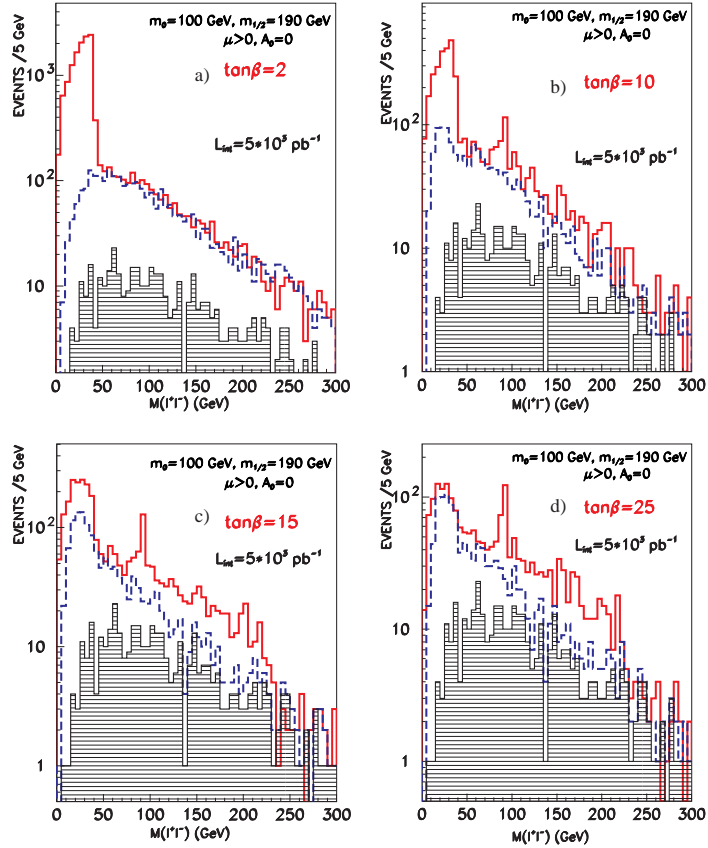

\dofig{0.85\textwidth}{denegrifig.ai}
\caption{Mass distributions for $e^+e^-+\mu^+\mu^-$ (solid) and
$e^\pm\mu^\mp$ (dotted) for various $\tan\beta$. \label{denegrifig}}
\end{figure}

	Decays into $\tau$'s can also be studied by comparing the
$e^+e^-+\mu^+\mu^-$ and $e^\pm\mu^\mp$ distributions.\cite{denegri}
As $\tan\beta$ is increased, the $\tau$ branching ratios become
larger, leading to more nearly equal distributions at low mass.  This
is illustrated in Figure~\ref{denegrifig}, which compares four values
of $\tan\beta$ for $m_0=100\,\GeV$, $\mhalf=190\,\GeV$, $A_0=0$, and
$\sgn\mu=+$. The mixing between gauginos and Higgsinos also increases,
leading to more heavy gaugino decays. These heavy gauginos can decay
both into $Z$'s and into $\tell^\pm\ell^\mp$, giving the higher-mass
structure in Figure~\ref{denegrifig}.

\section{GMSB Models}

	The phenomenology of GMSB models depends on the nature and
lifetime of the NLSP. Prompt $\lsp \to \tG\gamma$ or $\tell \to
\tG\ell$ decays give longer decay chains and hence more constraints.
A long-lived NLSP $\tell_R$ allows full reconstruction.  In this case
it is important to reconstruct muon-like tracks with $\beta<1$ and to
measure their mass with time of flight or $dE/dx$. A long-lived NLSP
$\lsp$ has signatures qualitatively like SUGRA.  In this case it is
important to search for rare $\lsp \to \tG\gamma$ decays, since these
measure the true SUSY breaking scale.  The ATLAS EM calorimeter
provides good angular measurements for photons and might be sensitive
to decay lengths up to $100\,\km$. All of these are discussed in much
more detail in Reference\,\citenum{TDR}. 

\section{Status and Outlook}

	The LHC produces many SUSY channels and a wide variety of
signatures involving combinations of $\etmiss$, jets, heavy flavors,
leptons, and $\tau$'s. The main background for these signatures is other
SUSY processes, not Standard Model ones. Given the many complex
possibilities, it is hard to draw general conclusions, but we reasonably
expect to make many precise measurements. It also seems likely that some
aspects of SUSY will be difficult to study at the LHC. The linear
collider studies should put particular emphasis on those aspects,
including heavy Higgs bosons, heavy Higgsinos, sneutrinos and sleptons
with $M_{\tell}>M_{\tchi_2^0}$, and dominant $\tau$ decays.

\medskip

	This work is supported in part by the U.S. Department of Energy
under contract DE-AC02-98CH10886.

\end{document}